\documentstyle[12pt]{article}

\begin{document}
\begin{titlepage}

\title { Persistent currents in coupled mesoscopic rings.}

\author{T. P. Pareek$^\dagger$ and A. M. Jayannavar$^\ddagger$.
\\thanks \\
Institute of Physics, Sachivalaya Marg, Bhubaneswar-751005, INDIA.}

\footnotetext[1]{e-mail:pareek@iopb.ernet.in}
\footnotetext[2]{e-mail:jayan@iopb.ernet.in}

\maketitle

\thispagestyle{empty}

\begin{abstract}

We have analysed the nature of persistent currents in open
coupled mesoscopic rings. Our system is comprised of two ideal
loops connected to an electron reservoir. We have obtained
analytical expressions for the persistent current densities in
two rings in the presence of a magnetic field. We show that the
known even-odd parity effects in isolated single loops have to
be generalised for the case of coupled rings. We also
show that when the two rings
have unequal circumferences, it is possible
to observe opposite currents (diamagnetic or paramagnetic) in
the two rings for a given Fermi level.

\end{abstract}

PACS NO : 75.20.-g, 72.10.Bg, 67.57.Hi

\end{titlepage}

\eject

\newpage
\hspace {0.5in}
{\bf I. Introduction}

\par It was predicted by B$\ddot u$ttiker et. al. [1]
that an equilibrium persistent current flows
in an ideal one dimensional mesoscopic ring
threaded by a magnetic flux $\phi$. Persistent current flows in a ring
as a response to magnetic field which
destroys the time reversal symmetry and is periodic in magnetic
flux , with a period $\phi_0$ , $\phi_0$ being the elementary flux
quanta ($\phi_0$=hc/e). At zero temperature the amplitude of
persistent current is given by e$v_f$/L , where $v_f$ is the Fermi
velocity and L is the circumference of the ring. For spinless
electrons persistent current can be either diamagnetic or
paramagnetic depending upon whether the total number of electrons
present is odd or even, respectively [2,3]. This behaviour of
persistent  current is also known as parity effect.
The existence of persistent currents in mesoscopic rings
has subsequently been confirmed by several experiments [4-6].

\par Persistent currents  occur in both open and isolated closed
systems [7-13]. A simple open system [7] is a metallic ring connected to an
electron
reservoir, characterized by chemical potential $\mu_1$.
Several new effects related to persistent currents can arise in
open systems which have no analog in closed or isolated systems.
Recently we have also shown [13-15] that large circulating currents can
arise in open mesoscopic rings in the presence of a transport
current ,in the absence of magnetic field. This is purely a
quantum effect and is related to the property of current magnification
in the loop.
\par So far theoretical treatments of persistent currents deal
with single rings(open and closed systems) threaded by magnetic
flux. Studies in a closed ring have been extended to include
finite temperature effects, multichannel rings, disorder,
spin-orbit coupling and electron-electron effects [2,3,16-14]. In our present
work we
study persistent currents in coupled
mesoscopic rings. Specifically, we consider two normal one
dimensional(single channel) coupled
rings connected through an one
dimensional ideal wire as shown in fig.(1).
The connecting lead makes contact with the left and the right ring
at junctions $J_1$ and $J_2$, respectively. This lead, in turn, is
connected to an external electron reservoir characterized by a chemical
potential $\mu$, via an another ideal lead making connection at
point X. The circumference
of the left and the right rings are $l_1$ and $l_2$ respectively. The
distances ${J_1}X$ ,${J_2}X$ are $l_3$, $l_4$,  respectively.
The electron reservoir acts as a source as well as a sink of electrons,
and by definition their is no phase relationship between the
absorbed and the emitted electrons. Electrons
emitted by the reservoir propagate along the lead and are
partially reflected by the junction point X and are partially
transmitted along the loop via repeated scatterings at junction
points X, $J_1$, and $J_2$. Electrons in the loops will
eventually reach the reservoir after some time delay. This gives
rise to finite life time broadening for the electron states
of the coupled rings. Scattering processes in the coupled rings
are elastic. Only the reservoir acts as an inelastic scatterer.
There is complete spatial separation between the sources of
elastic and inelastic scattering. Our present analysis concerns
non-interacting spinless electrons. In the presence of an
external uniform magnetic field B, The magnetic flux through the
left and the right rings are given by $\alpha_1=B{l_1}^2/4\pi$
and $\alpha_2=B{l_2}^2/4\pi$, respectively; $\alpha_1$
and $\alpha_2$ are not independent quantities as the magnetic
flux in both the rings arise from the same applied uniform
magnetic field B. We have obtained analytical expressions for
the persistent current densities in both the rings in the
presence of a magnetic field. We show that persistent currents
in the two rings are very sensitive to the geometric features ( such
as lengths $l_1,l_2,l_3$ and $l_4$) of the system. Even though
we have obtained an analytical expression for the general case,
we restrict ourselves to a case where $l_3=l_4$. When the rings
are of the same size the magnitude and sign of the persistent
currents are same in both the rings (due to symmetry). We
observe that if the distance between the rings $l=(l_3+l_4)$ is
much larger than the circumference of the two identical rings
the known even odd parity effect (known for isolated rings)
breaks down [20]. In the second case we have considered rings of
unequal circumferences. In such a situation at a given Fermi
level it is possible to observe diamagnetic current in one of the rings
and simultaneously paramagnetic current in the other ring.
In sec.II we give a brief account of theoretical treatment and
sec.III is devoted to results and conclusions.

\noindent {\bf .II Theoretical treatment}
   \par In this section we derive expressions for persistent current
   in the left and the right rings for the general case when $l_3$$\neq$$l_4$.
   For this we consider a noninteracting electron system.
   Our system is considered as a free electron network, i.e., the
quantum potential V throughout the network is assumed to be
identically zero. The scattering arises solely due to junctions
(or geometric scattering) at $J_1$, $J_2$ and X.
   For scattering at the junctions we do not assume any specific
   form of scattering matrix , instead the junction scattering matrix
   follows from principles of quantum mechanics. We use the
   Griffiths boundary condition (current conservation) and the single
   valuedness of wave function at the junctions [21]. Following
exactly the same procedure as the earlier ones [11-15] one can readily derive
analytical expressions for the persistent current
densities (dJ/dk) (i.e. persistent current density in the small wave
vector interval k and k+dk) in the left ($dJ_{L}/dk$)
and right ($dJ_{R}/dk$) loops [7] and are given by

\begin{eqnarray}
dJ_{L}/dk=- (ek\hbar/2\pi m)256 Sin [\alpha] Sin [kl_1] \left\{ Sin [k( l_2 -
l_4 )] \right .\nonumber\\
\left.  - 3 Sin [k( l_2 + l_4)] - 2 Sin [\beta - kl_4]
      + 2 Sin [\beta + kl_4] \right\}^2/({\Omega_1}^2+{\Omega_2}^2),
\end{eqnarray}

\begin{eqnarray}
dJ_{R}/dk=-(ek\hbar/2\pi m) 256 Sin [\beta] Sin [kl_2] \left\{ Sin [k( l_1 -
l_3 )] \right. \nonumber\\
\left.  - 3 Sin [k( l_1 + l_3)] - 2 Sin [\alpha - kl_3] +
       2 Sin [\alpha + kl_3] \right\}^2/({\Omega_1}^2+{\Omega_2}^2).
\end{eqnarray}

\noindent Here $\Omega_1$ and $\Omega_2$ are given by

\newpage
\begin{eqnarray}
\Omega_1=2 \left\{ -3 Cos (k\left( l_1 - l_2 - l_3 - l_4 \right) )
    - Cos (k\left( l_1 + l_2 - l_3 - l_4 \right) )
    + 9 Cos (k\left( l_1 - l_2 + l_3 - l_4 \right) )  \right . \nonumber\\
\left. + 3 Cos (k\left( l_1 + l_2 + l_3 - l_4 \right) )
     + Cos (k\left( l_1 - l_2 - l_3 + l_4 \right) )
     + 3 Cos (k\left( l_1 + l_2 - l_3 + l_4 \right) )  \right .\nonumber\\
\left. - 3 Cos (k\left( l_1 - l_2 + l_3 + l_4 \right) )
     - 9 Cos (k\left( l_1 + l_2 + l_3 + l_4 \right) ) - \right .\nonumber\\
\left. 4 Cos (\alpha - \beta - kl_3 - kl_4) - 4 Cos (\alpha + \beta - kl_3 -
kl_4) +
      6 Cos (\beta - kl_1 - kl_3 - kl_4) \right .\nonumber\\
\left.     + 2 Cos (\beta + kl_1 - kl_3 - kl_4)
  + 6 Cos (\alpha - kl_2 - kl_3 - kl_4) + 2 Cos (\alpha + kl_2 - kl_3 -
kl_4)\right .\nonumber\\
\left. + 4 Cos (\alpha - \beta + kl_3 - kl_4) + 4 Cos (\alpha + \beta + kl_3 -
kl_4)
       - 2 Cos (\beta - kl_1 + kl_3 - kl_4)\right .\nonumber\\
\left. - 6 Cos (\beta + kl_1 + kl_3 - kl_4)
      - 6 Cos (\alpha - kl_2 + kl_3 - kl_4) - 2 Cos (\alpha + kl_2 + kl_3 -
kl_4) \right .\nonumber\\
\left. + 4 Cos (\alpha - \beta - kl_3 + kl_4) + 4 Cos (\alpha + \beta - kl_3 +
kl_4)
      - 6 Cos (\beta - kl_1 - kl_3 + kl_4) \right. \nonumber\\
\left. - 2 Cos (\beta + kl_1 - kl_3 + kl_4)
       - 2 Cos (\alpha - kl_2 - kl_3 + kl_4) - 6 Cos (\alpha + kl_2 - kl_3 +
kl_4) \right .\nonumber\\
\left. - 4 Cos (\alpha - \beta + kl_3 + kl_4) - 4 Cos (\alpha + \beta + kl_3 +
kl_4)
       + 2 Cos (\beta - kl_1 + kl_3 + kl_4) \right. \nonumber\\
\left. + 6 Cos (\beta + kl_1 + kl_3 + kl_4)
       + 2 Cos (\alpha - kl_2 + kl_3 + kl_4) + 6 Cos (\alpha + kl_2 + kl_3 +
kl_4) \right\},
\end{eqnarray}

\begin{eqnarray}
\Omega_2=4 \left\{ -3 Sin (k\left( l_1 - l_2 - l_3 - l_4 \right) ) -
    Sin (k\left( l_1 + l_2 - l_3 - l_4 \right) ) + \right . \nonumber\\
\left .    3 Sin (k\left( l_1 - l_2 + l_3 + l_4 \right) ) +
    9 Sin (k\left( l_1 + l_2 + l_3 + l_4 \right) ) -  \right . \nonumber\\
\left .    4 Sin (\alpha - \beta - kl_3 - kl_4) - 4 Sin (\alpha + \beta - kl_3
- kl_4) +
    6 Sin (\beta - kl_1 - kl_3 - kl_4) +  \right . \nonumber\\
\left .    2 Sin (\beta + kl_1 - kl_3 - kl_4) +
    6 Sin (\alpha - kl_2 - kl_3 - kl_4) + 2 Sin (\alpha + kl_2 - kl_3 - kl_4) +
 \right . \nonumber\\
\left .    4 Sin (\alpha - \beta + kl_3 + kl_4) + 4 Sin (\alpha + \beta + kl_3
+ kl_4) -
    2 Sin (\beta - kl_1 + kl_3 + kl_4) -  \right . \nonumber\\
\left .    6 Sin (\beta + kl_1 + kl_3 + kl_4) -
    2 Sin (\alpha - kl_2 + kl_3 + kl_4) - 6 Sin (\alpha + kl_2 + kl_3 + kl_4)
\right\},
\end{eqnarray}

\noindent where $\alpha=2\pi\alpha_1/\phi_0$ and
$\beta=2\pi\alpha_2/\phi_0$,$\phi_0=hc/e$ the elementary flux quanta.
The wavevector of an electron is denoted by k and is related to
the energy E of an electron by a simple relation E=$\hbar^2 k^2/2m$.
Since we are considering the case wherein
the magnetic field B is due to the same source and consequently
the flux ($\alpha$ and $\beta$ are written in a dimensionless
form)   piercing through the two loops are related by the
following relation (i.e., $\alpha$ and $\beta$ are dependent
variables).
     \begin{equation}
    \alpha=(\left ({l_{2}}^2/{l_{1}}^2)*\beta\right).
   \end{equation}

\par For the above case from equation (1)-(5) one can
readily verify that persistent current densities are
antisymmetric in B or the persistent currents in two loops
change sign on the reversal of magnetic field (B $\rightarrow$ $-$B).
Henceforth we rescale the current densities in the
dimensionless form and denote $dj_{L}=(dJ_{L}/dk)(2m\pi/\hbar ek)$ and
$dj_{R}=(dJ_{R}/dk)(2m\pi/\hbar ek)$. We have also rescaled all the
lengths with respect to the length $l_1$ of the left hand loop.
The wave vector is written in the dimensionless form as k$l_1$.

\noindent {\bf .III Results and Discussions}
\par We would like to point out that our expression for the
persistent current densities obtained in equation (1) and (2)
are quite general and valid even for the case, where, the flux
enclosed by two rings $\alpha_1$ and $\alpha_2$ are independent
variables. This case corresponds to a situation in which the
enclosed magnetic flux in the left and right rings may arise
respectively from two independent magnetic field sources.
However, in our present detailed analysis we have not considered
this case. If the two rings are identical ($l_1=l_2$) we notice
that the magnitude of the persistent current densities in the
left and the right rings are unequal. This follows from the fact
that there is a asymmetry in the system. This asymmetry arises
because of the junction scattering point X, which is not placed
at a symmetrical position with respect to the position of the
two rings ($l_3\neq l_4$). Henceforth we restrict our discussion
further to the case $l_3 = l_4$ (symmetrical situation). For
this special case, when $l_1 = l_2$, the magnitude and the
direction of the persistent current are same in both the rings.

\par In fig.2 we have plotted persistent current density
$dj_{L}$ as a function of dimensionless wave vector k$l_1$ for
a fixed value of $l_2/l_1$=1,
$l_3/l_1$=$l_4/l_1$=0.5 , and $\alpha$=0.5. Since in this particular case
the system is symmetric about the junction X , we expect that current in
the left or the right ring will be same. As one varies $kl_1$ the persistent
currents
oscillate between diamagnetic and paramagnetic behaviour.
In our problem the coupled rings are connected to a
reservoir, which, in turn, leads to finite life time broadening of
the electron states in the system and as a consequence the
persistent current shows a broadened feature as a function of
k$l_1$ compared to an isolated system. The amplitude extrema in
persistent current occur approximately at the values of
k$l_1$=2$\pi(n+\alpha_1/\phi_0)$ , where n=0,$\pm$1,$\pm$2,...etc., which
correspond to the allowed states in a single isolated loop of length $l_1$. The
observed small deviation from values of k$l_1$ for isolated ring
follows from the fact that there is a coupling between the rings
and additional scatterings at $J_1 , J_2$ and X.

\par In fig.(3) we have plotted persistent current as
a function of $\alpha$ for a fixed value of $kl_1=6.0$. Other
parameters being the same as used for fig.(2).
We notice that results obtained in fig.2 and fig.3. are qualitatively same as
one observes in a single loop of length $l_1$ connected to an
electron reservoir [7]. It is also to be noted that  the simple
periodicity observed in fig.(2) and fig.(3) is due to the fact that
all lengths ($l_1, l_2, l_3$, and $l_4$) are simple rational
multiples of each other, otherwise  we would have obtained a
complicated structure in the behaviour of persistent current as
a function of k$l_1$ as well as $\alpha$.

\par From now on we discuss the case when the length of the connecting
lead ($l_3+l_4$) is much larger than the circumference $l_1$ of
the loops. We have taken both the loops to be of equal circumference.
We show that in this simple case the even-odd parity effect known
for isolated rings breaks down and the parity effect gets
modified in accordance with the length ratio $(l_3+l_4)/l_1$. In the
absence of magnetic field, an isolated single loop has eigenstates
corresponding to wave vector k =$2\pi n/l_1$, whereas isolated
connecting wire of length $l_3+l_4$ has eigen states with k =
$n\pi/(l_3+l_4)$ (n=0,$\pm$1,$\pm$2,...etc.). Therefore for the length
$(l_3+l_4) > l_1$ energy levels in the isolated lead are closely
spaced than the energy levels in the loop. These closely spaced
energy levels, leak into the loops (hybridized with the states
within the loop) in a connected ring system and consequently
additional quasi bound states arise which have energies lying between the
states of the isolated ring. Naturally the energies of these new
states of the coupled system will be shifted from either of those of the
separate lead and the ring due to the coupling(perturbation). In the
presence of magnetic field such an additional state contributes to
the persistent current diamagnetically or paramagnetically
depending on whether it is near respectively to a diamagnetic or
paramagnetic state of the isolated loop (in the presence of a
magnetic field). These states basically owe their existence to the
resonant states in the isolated lead and their contribution to the
magnitude of persistent current is small compared to the
contribution of persistent current from the states near the
resonant states of the isolated loops. Thus a situation can arise
a system of coupled loops (with $(l_3+l_4) >> l_1$) such that first
$N_1$ states are diamagnetic and the next $N_1$ states will be
paramagnetic (2$N_1$ is the number of resonant states, of the
lead, lying between the two successive levels of the isolated
ring)  and process repeats as we go to higher states. In
a single isolated loop, for spinless electrons, it is well known[2,3]
that current in a loop is diamagnetic or paramgnetic depending
on whether the number of particles is odd or even, respectively(even-odd
parity effect). Now for coupled mesoscopic rings this simple
even-odd parity effect gets altered and instead we have first
$N_1$ levels contributing to a diamagnetic current but the next $N_1$
levels contribute a paramagnetic current, where $N_1$ depends on the ratio
$(l_3+l_4)/l_1$. This is true for the case of two identical
loops. Parity effect will have different meaning if $l_1\neq l_2$
(non identical loops). For this case the underlying concepts will become a
little
complicated as we have to discuss parity effects in the left and
right loops separately as they carry different currents for any
given state, which will be discussed below. In fig.(4) we have
plotted the persistent current $dj$ as a function of k$l_1$, for
the case when $(l_3+l_4)/l_1$=2,$l_2/l_1$=1
and for a fixed value of $\alpha=1.2$.
For this situation we have two additional states of the
connecting lead (lying between eigen states of the isolated loops),
which leak into the loops. We clearly observe from fig.(4) that as we
vary k$l_1$ we get the first two peaks which are diamagnetic and
the later two peaks are paramagnetic and the sequence repeats.

\par In our
problem we basically solve a scattering problem wherein electrons
are injected in the system from the reservoir which get
reflected back to the reservoir. From a scattering matrix
structure one can get the information about quasi bound states.
This can be achieved by looking at the poles in a complex $kl_1$ plane
of the complex reflection amplitude. The real
part of the poles (R) gives the wave vector values of the resonant
states, whereas the imaginary part gives the information about the
lifetime of these states. In fig.(5) we have plotted the real
part R of these complex poles as a function of
$\alpha$. All the parameters used here are the same as in
fig.(4). We clearly observe that additional states (in
the present case 2) appears within the intervals of k$l_1$ values of
isolated loops. Moreover, one can readily notice that the first
two resonant states carry diamagnetic current (as their
slopes with respect to the magnetic flux are positive [2,3]) and the next two
carry paramagnetic currents and so on. As expected on the general
grounds values of R are periodic in flux $\alpha_1$ with a
period $\phi_0$. In fig.(6) we have plotted dj versus $kl_1$ for
the case $(l_3+l_4)/l_1=10.0 $ and for a fixed value of
$\alpha=1.2$ while $l_2/l_1=1.0$. It
is clear from this figure that the first six states carry
diamagnetic current, next six states carry paramagnetic
current and so on. The fig.(4) and fig.(6) clearly indicate that
the known even-odd parity effect in an isolated ideal ring
breaks down for a system of coupled rings and moreover the
emergence of new parity effect as discussed above is sensitive
to the length ratio $(l_3+l_4)/l_1$.

\par We further consider the case for which the loops are not
identical, in that their circumferences are different. In such a
situation one has to discuss the persistent currents in the right
and the left loops separately. Consider a situation where $l_2 >
l_1$. Naturally resonant states in the right loop are more
closely spaced than those in the left loop. There will be mixing between
these states due to the coupling. However, it is possible that
as one varies wave vector k$l_1$ persistent current in the right
loop oscillates between diamagnetic and paramagnetic behaviour
much more rapidly than the persistent current in the left hand
loop, i.e., in a given interval of k$l_1$, persistent current
does not change sign for the case of left hand loop whereas in
the same interval persistent current in the right hand loop changes
sign several times. We can have a situation where for a given
state (k$l_1$) current in the left and right loops have either the same sign or
different (i.e., current in left loop are diamagnetic where as
current in the right loop is paramagnetic). This is illustrated
in fig.(7), where .... lines and ------ lines indicate
persistent current in the right(d$j_{R}$) and the left
(d$j_{L}$) loop, respectively. For the
above case we have taken $l_2/l_1=4$ and $(l_3+l_4)/l_1=1$. In
fig.(8) and fig.(9) we have plotted persistent currents as a
function of $\alpha$ for a fixed value of $kl_1=2.2$ for the right
and the left loop respectively. The other parameters are same as
those used in fig.(7). From fig.(8) and fig.(9) we notice that
$dj_{L}$ and $dj_{R}$ are periodic in $\alpha_1$ with a period
$\phi_0$. We would like to mention that this is so because for our
case we have considered a commensurate ratio $l_2/l_1=4$. In
general if we choose the ratio to be incommensurate (or
irrational) $dj_{L}$ or $dj_{R}$ will have much larger value of
the period with respect to $\alpha$. It should be kept in
mind that as one varies $\alpha_1$ (the flux through the left
ring) by $\phi_0$, the flux through the right ring ($\alpha_2$)
changes by an amount 16$\phi_0$ (since $l_2=4l_1$). It follows
from this that as $\alpha$ is varied from 0 to $2\pi$ the
persistent current density in the left loop changes sign once
while the persistent current density in the right loop changes
the sign 16 times.

\par It is well known that for a
simple case of isolated single loop (or a single hole in the
sample) the persistent current carried by the nth state of energy $E_n$
is given by $I_n=-(1/c)\partial \epsilon_n / \partial \phi$, where $\phi$ is
the
flux piercing through the loop (or hole). In the present case
of multiply connected nonidentical rings one cannot infer the
values of persistent current in the individual rings from the above
definition. To calculate persistent current in the presence of
magnetic field in each loop of the system of coupled rings one has
to calculate quantum mechanical wave function in each ring explicitly
and from that one can calculate the currents.

\par In our analysis throughout we have discussed the persistent
current densities dj in the small wave vector interval k and
k+dk. However, experimentally it is the total persistent current
generated by all the conducting
electrons in the system that can be observed.
This can be calculated by integrating the
persistent current densities up to the Fermi wave vector $k_f$ using
eqns.(1) and (2). In conclusion , we have studied the nature of persistent
currents
in open mesoscopic coupled ring system in presence of magnetic
field. Throughout we have considered simple commensurate ratios
of $l_1/l_2$ and $(l_3+l_4)/l_1$. For coupled identical rings
one observes different parity effects. The parity effect depends
on the ratio $(l_3+l_4)/l_1$ of the connecting lead length to
the circumference of the rings. In the case of non identical
loops, for a given state, it is possible that persistent current
in one loop is diamagnetic whereas in the other it may be
paramagnetic or diamagnetic. Moreover all these effects are very
sensitive to the length ratio involved in the system as the
problem is inherently quantum mechanical in nature, where
interference effects dominate.

\vfill

\eject

\vfill
\eject
{\bf Figure captions}

Fig. 1. Two metal loops connected to an electron
reservoir with chemical potential $\mu_1$.

Fig. 2. Plot of circulating current versus k$l_1$ for a fixed
value of $\alpha$=0.2. For this case $l_2/l_1=1.0$,
$l_3/l_1$=$l_4/l_1$=0.5.

Fig. 3. Plot of circulating current versus $\alpha$ for a fixed
value of $kl_1=6.0$. For this case $l_2/l_1=1.0$,
$l_3/l_1$=$l_4/l_1$=0.5.

Fig. 4. Plot of circulating current versus k$l_1$ for a fixed
value of $\alpha$=1.2. For this case  $l_2/l_1=1.0$,
$l_3/l_1$=$l_4/l_1$=1.0.

Fig. 5. The plot of real part R of the complex poles in the
$kl_1$ plane of the reflection amplitude as a function of $\alpha$
for $l_2/l_1=1.0$,
$l_3/l_1$=$l_4/l_1$=1.0.

Fig. 6. Plot of circulating current versus k$l_1$ for a fixed
value of $\alpha$=1.2. For this case $l_2/l_1=1.0$,
$l_3/l_1$=$l_4/l_1$=5.0.

Fig. 7. The persistent current as a function of
$kl_1$ in the left loop (solid
line) and the right loop (dashed lines) for a fixed value of
$\alpha=1.2$. For this case
$l_2/l_1=4$ and $l_3/l_1=l_4/l_1$=0.5.

Fig. 8. Plot of persistent current in the left loop as a
function of $\alpha$ for a fixed value of $kl_1=2.2$.
For this case $l_2/l_1=4$ and $l_3/l_1=l_4/l_1$=0.5.

Fig. 9. Plot of persistent current in the right loop as a
function of $\alpha$ for a fixed value of $kl_1=2.2$.
For this case
$l_2/l_1=4$ and $l_3/l_1=l_4/l_1$=0.5.

\vfill
\eject
\end{document}